\documentstyle[preprint,pra,aps]{revtex}
\begin{document}
\draft
\title{First order metamagnetic transition in CeFe$_2$ based pseudobinary 
alloys}
\author{Meghmalhar Manekar, Sujeet Chaudhary, M. K. Chattopadhyay, Kanwal
Jeet Singh,\\ S. B. Roy and P. Chaddah}
\address{Low Temperature Physics Laboratory,\\
Centre for Advanced Technology,\\ Indore 452013, India}
\date{\today}
\maketitle
\begin{abstract}
We present results of dc magnetisation study showing that the 
low temperature antiferromagnetic state in various CeFe$_2$-based
pseudobinary alloys  can be transformed into ferromagnetic state 
through a magnetic field induced phase transition. 
We highlight the presence of hysteresis and phase coexistence
across this metamagnetic  transition and argue that the observed 
phase transition is of first order in nature.
\end{abstract}                          

\section{introduction}
The interesting magnetic properties of the C15-Laves phase compound CeFe$_2$ 
\cite{1,2,3,4,5,6} and its 
pseudobinaries \cite{7,8,9,10,11,12,13,14,15,16,17,18,19,20} have been drawing 
almost continuous attention during last twenty years. Most of these studies
are mainly focussed on the proper understanding  of 
the magnetic ground states of the parent as well as 
the pseudobinary compounds, and less emphasis
is given on the exact nature of the phase transitions. 
In a recent study \cite{21} we have addressed this latter question in 
the context of double magnetic transitions in  CeFe$_2$ based 
pseudobinary systems. With the temperature
dependent ac-susceptibility measurements we have shown that while 
the paramagnetic to ferromagnetic transition (as a function of decreasing 
temperature) is a second order phase transition, the lower temperature 
ferromagnetic to antiferromagnetic transition carries the signature of a first order
phase transition. The presence of thermal hysteresis and phase coexistence 
across this ferromagnetic to antiferromagnetic transition was highlighted \cite{21}. 
With the existing information \cite{13,15,17,18,19,20} that the lower temperature 
antiferromagnetic state can be reverted back to the ferromagnetic state
with the application of an external magnetic field, the 
question naturally arises - what is the nature of this field induced 
antiferromagnetic to ferromagnetic transition ? In this paper we address this
question in details through our dc magnetisation measurements. We shall argue 
that this metamagnetic transition is of first order in nature.

\section{Experimental}
We have used in the present study the same two samples 
-- Ce(Fe,5\%Ir)$_2$ and Ce(Fe,7\%Ru)$_2$-- used in our earlier 
measurements \cite{21}. Dc magnetisation is measured using a commercial SQUID
magnetometer (Quantum Design-MPMS5). We have used a scan length of 2cm, with 
each scan containing 32 data points. The 2cm scan length is used to ensure 
minimum sample movement in the inhomogeneous magnetic field of the 
superconducting magnet. This magnetic field inhomogeneity in a 2 cm scan 
is about 1 Oe in an applied field of 20 kOe \cite{22}.  

\section{Results and discussion}
In Fig. 1 we present magnetisation (M) versus tempearture (T) plots for 
the Ce(Fe,5\%Ir)$_2$ and Ce(Fe,7\%Ru)$_2$ samples obtained in an 
applied field of 100 Oe. The double 
magnetic transitions are clearly visible and the transition temperatures are 
well in accord with those obtained earlier in ac-suceptibility measurements
\cite{21}. The data shown in Fig. 1 are obtained while warming up 
unidirectionally from  low temperatures after zero field cooling. We have also obtained data while 
cooling and a distinct thermal hysteresis of width 5 K is observed across 
the ferromagnetic to antiferromagnetic transition. This is to be contrasted with the
relatively smaller thermal hysteresis of 2 K obtained earlier in the 
ac-suceptibility measurements\cite{21}. We note that a smaller hysteresis is expected in an ac
measurement because the ac field assists a metastable to stable state transformation.Our SQUID 
magnetometer cannot, however, monitor possible temperature lags between the sample and the 
sensor - as we were able to do in our ac measurements \cite{21}. For this reason we shall 
not emphasize thermal hysteresis in this report. 

In Fig. 2 we present isothermal magnetisation (M) versus field (H) plots 
for Ce(Fe,5\%Ir)$_2$ at various temperatures obtained after 
zero field cooling  the sample from tempearture above the Curie 
tempeature (T$_C \approx$185K). Above 130 K the M-H plot shows the 
typical behaviour
of a ferromagnet, reaching technical saturation by 10 kOe at 150K. Below 80K, 
although there is a small non-linearity in the low field ( H $<$ 5 kOe) regime 
the character of the M-H plots are drastically different. This we attribute to
the antiferromagnetic nature of the low temperature phase. The M-H behaviour in
the T regime 130K $> T >$ 80K, however, is quite interesting (see Fig. 2) .
While in the low field regime there is a distinct deviation from characteristics of 
the higher temperature ferromagnetic state
, a sharp rise in M occurs in the high H regime 
indicative of a field induced ferromagnetic transition or a
metamagnetic transition.  We mark the onset field of this metamagnetic
transition as H$_M$. H$_M$ is estimated as the field at which the M-H curve changes curvature from convex to concave,  i.e. where dM/dH shows a minimum.  With the decrease in T the 
value of H$_M$ increases and goes beyond the range of existing field 
strength (55 kOe) in our SQUID magnetometer by 60K.    

We have obtained qualitatively similar results from the isothermal field dependence of magnetisation at various temperatures for the Ce(Fe,7\%Ru)$_2$ 
sample measured after zero field cooling the sample from temperature above Curie temperature 
(T$_C \approx$165K). These M-H plots are shown in Fig. 3.

We shall now focus on this metamagnetic transition, and look for signatures
typically associated with a first order phase transition, namely hysteresis
and phase coexistence.The control variable inducing the transition is magnetic field.
 We shall work in the temperature regime, 
80K$\leq T \leq$130K for
Ce(Fe,5\%Ir)$_2$ and 90K$\leq T \leq$130K for Ce(Fe,7\%Ru)$_2$, 
so that the metamagnetic 
transition remains clearly visible within the upper limit of our magnetic field
range. We present in Fig.4-5 M-H curves obtained in the ascending and descending
field cycles for both 
Ce(Fe,5\%Ir)$_2$ and Ce(Fe,7\%Ru)$_2$, showing distinct hysteresis associated
with the metamagnetic transition. A sharp rise in magnetisation accompanied 
by hysteresis is traditionally attributed to the
first order magnetic process \cite{24}. However, it can still be argued that the
observed hysteresis may be the intrinsic property of the field induced 
ferromagnetic state and originates from the domain wall pinning and/or 
freezing of domain rotation \cite{25}. To negate these possibilities 
 we have measured carefully the M-H curves for both the samples 
in the temperature regime 
where the ground state is ferromagnetic at all H values. M-H curves in the
ferromagnetic regime show negligibly small hysteresis with the 
coercivity field (H$_C$) of the order of 5 Oe. 
To show the contrast of 
the hysteresis intrinsic to the ferromagnetic regime of these samples with the
hysteresis associated with the metamagnetic transition, in Fig. 6 we present
M-H curves for Ce(Fe,7\%Ru)$_2$, measured at T=130 K and 120K.
(The ferro- to antiferromagnetic transition temperature for this sample takes
place approximately at 127 K (see Fig. 1 (b)). At T = 120K,
the onset of the metamagnetic transtion takes place at a relatively small 
value of H$_M\approx$1.5kOe. The hysteresis associated with 
the metamagnetic 
transition shows up as a distinct bubble in the field regime 
4kOe$<H<$30kOe, with
relative reversibility in the field regime above (see fig.5) and below. This is to be contrasted with the M-H
curve at T=130K which is quite reversible (in the same scale) 
for all H values of measurement (see
Fig.6). We have also checked the field history dependence of the hysteresis
associated with the metamagnetic transition by cycling the 
field isothermally (after initial zero field cooling) 
more than once between zero and the maximum applied field (50 kOe). Unlike in
the case of ferromagnetic hysteresis loops, no virgin curve is 
observed here, and the obtained hysteresis loop in the first field cycle is reproduced in all the 
subsequent cycles.

While the hysteresis associated with the 
ferromagnetic state is relatively small and does not change 
much in the temperature regime 130K$<T<$160K ( measured but not shown here for
the sake of conciseness ), the hysteresis associated with the
metamagnetic transition is observed below 130K and grows relatively
rapidly with the decrease in T. It should be noted here that 
at T = 80K for  Ce(Fe,5\%Ir)$_2$ and at T = 90K for 
Ce(Fe,7\%Ru)$_2$ the formation
of the higher field ferromagnetic state is probably not completed by 50 kOe,
and accordingly the magnetisation and the 
associated hysteresis has not reached its saturation (see Fig. 4 and 5).

After establishing the hysteretic nature of the metamagnetic 
transition, we
shall now look for the phase coexistence in the transition region. To study
the phase co-existence we use the technique of minor hysteresis loops (MHLs)
\cite{21,26}. We shall first define the  hysteretic M-H curve obtained by isothermal
field cycling between 0 and H$_{max}$ = 50 kOe as the 
`envelope curve'. The field increasing curve (0 to 50 kOe) corresponds to the antiferromagnetic 
phase transforming to the ferromagnetic phase,with the antiferromagnetic phase persisting as a 
metastable phase over some field region. Similarly the field decreasing (50 kOe to 0) curve
corresponds to the ferromagnetic phase transforming to the antiferromagnetic
phase with the ferromagnetic phase persisting as a metastable phase over some field region.
 We can now generate an MHL during the ascending field cycle 
i.e. start increasing H from the lower field reversible (antiferromagnetic)
regime and  then reverse the direction of change of H before reaching the higher field
saturation magnetisation regime. We can also produce an 
MHL in the
descending H cycle i.e. start decreasing H from the saturation magnetisation
regime and reverse the direction of change of H before reaching the low H antiferromagnetic
regime. We show in Fig. 7 and 8 examples of these MHLs in 
 Ce(Fe,5\%Ir)$_2$ and Ce(Fe,7\%Ru)$_2$ samples.

At field values close to H$_M$ on the ascending field cycle, the high field ferromagnetic phase is not expected to be formed in sufficient quantitiy, and complete transformation occurs only at much higher fields. When we initiate an MHL from a field close to H$_M$, this partially formed ferromagnetic phase `supercools' and persists as a metastable phase. We see only small amount of hysteresis as H is lowered from field values close to H$_M$ (see inset of fig. 7(a) and 8(a)). At fields well above H$_M$ a much larger fraction (close to 100\%) of the  sample has transformed to the ferromagnetic phase. When we now lower H and initiate an MHL, the entire sample `supercools' in the ferromagnetic phase, and the hysteresis observed should be much larger. This is brought out in figures 7 and 8 where MHLs initiated from  field values well inside the hysteretic regime
coincide with the upper envelope curve, 
indicating that the 
high field ferromagnetic phase has formed in sufficient quantities. To show
further evidence of supercooling of the high field phase, in fig. 9 we show 
MHLs at H = 2kOe drawn from the lower envelope curve and at H = 1.6kOe drawn
from the upper envelope curve of the Ce(Fe,7\%Ru)$_2$ sample at T = 110K. These
field values are chosen such that the lower and upper envelope curves 
have the same magnetisation value. Note that 
H= 2 kOe is lower than the estimated H$_M\approx$4 kOe and accordingly the MHL 
drawn from
the lower envelope curve shows almost no irreversibility.The high field ferromagnetic phase is
 thus not yet formed. On the other hand
the MHL drawn from the upper envelope curve at a lower field value of 1.6 kOe
shows distinct irreversibility. This clearly shows that the high field phase
persists (as a metastable `supercooled' phase) in this field regime 
in the descending field cycle. Similar results
exist for the Ce(Fe,5\%Ir)$_2$ sample also, but not shown here for the sake of
conciseness.

It is to be noted here that the low field magnetic response of the 
low temperature (supposedly) antiferromagnetic state for both the 
samples is quite non-linear in
nature (see Fig. 2 and 3). This behaviour definitely points out the presence of some
ferromagnetic correlation in this low temperature phase. While it was pointed out earlier
that such a behaviour probably arose due to an impurity ferromagnetic 
phase \cite{20}, an intrinsic origin of such a behaviour cannot be 
ruled out entirely \cite{19}. 
Careful microscopic measurements ( like neutron diffraction and/or 
Mossbauer measurements ) are now necessary to resolve this 
problem of the CeFe$_2$-based pseudobinaries.

\section{Conclusion}
Summarising our results we say that the field induced 
antiferromagnetic to ferromagnetic transition in 
Ce(Fe,5\%Ir)$_2$ and Ce(Fe,7\%Ru)$_2$ samples is accompanied by field hysteresis
as well as phase coexistence. These are the typical characteristics of a first
order transition. Hence we argue that the observed metamagnetic transition in
these CeFe$_2$ based pseudobinaries is a first order transition. The present 
study compliments our earlier study of temperature variation in the same systems,
and establishes the existence of a first order 
ferromagnetic to antiferromagnetic phase transition in the CeFe$_2$ based 
systems in a more general H-T plane.

\begin{figure}
\caption{Magnetisation versus temperature plots for (a) Ce(Fe,5\%Ir)$_2$ and 
(b) Ce(Fe,7\%Ru)$_2$ }
\end{figure}
\begin{figure}
\caption{Magnetisation versus field plots for Ce(Fe,5\%Ir)$_2$ at various temperatures.
 The arrows mark the onset field H$_M$ of the metamagnetic transition.
 The line for the M-H curve at T=100K serves as a guide to the eye.}
\end{figure}
\begin{figure}
\caption{Magnetisation versus field plots for Ce(Fe,5\%Ir)$_2$ at various 
temperatures. The onset field H$_M$ of metamagnetic transition is 
marked by arrows.The line for the M-H curve at T = 100K 
serves as a guide to the eye.}
\end{figure}
\begin{figure}
\caption{M-H curves for Ce(Fe,5\%Ir)$_2$ showing hysteresis associated with the metamagnetic 
transition. The arrows show the direction of field change.}
\end{figure}
\begin{figure}
\caption{M-H curves for Ce(Fe,7\%Ru)$_2$ showing hysteresis associated with the 
metamagnetic transition. The arrows show the direction of field change.}
\end{figure}
\begin{figure}
\caption{M-H curves for Ce(Fe,7\%Ru)$_2$ at temperatures 120K and 130K. See text for details.}
\end{figure}
\begin{figure}
\caption{Minor hysteresis loops generated during (a) ascending field cycle for Ce(Fe,5\%Ir)$_2$
at H = 20kOe (open triangle), H = 26kOe (open circle), H = 35kOe (solid triangle) 
and H = 42.5kOe (solid circle) . 
(b) descending field cycle for Ce(Fe,5\%Ir)$_2$ at H = 40kOe (open square) and 
H = 45kOe (solid triangle). The envelope curve is represented by solid squares. The measurements were done at T = 85K. Inset shows the expanded view.}
\end{figure}
\begin{figure}
\caption{Minor hysteresis loops generated during (a) ascending field cycle for Ce(Fe,7\%Ru)$_2$
at H = 4kOe (open triangle), H = 10kOe (open circle) and H = 20kOe (open square). 
(b) descending field cycle for Ce(Fe,7\%Ru)$_2$ at H = 8kOe (open square) and H = 22kOe (open circle). 
The envelope curve is represented by solid squares. The measurements were done at T = 110K. Inset shows the expanded view.}
\end{figure}
\begin{figure}
\caption{Comparison of minor hysteresis loops generated during ascending field cycle (H = 2kOe - open circle)
and descending field cycle (H = 1.6kOe - open triangle) at approximately same value of 
magnetisation for Ce(Fe,7\%Ru)$_2$ at T=110K. The envelope curve is represented by solid squares.See text for details.}
\end{figure}

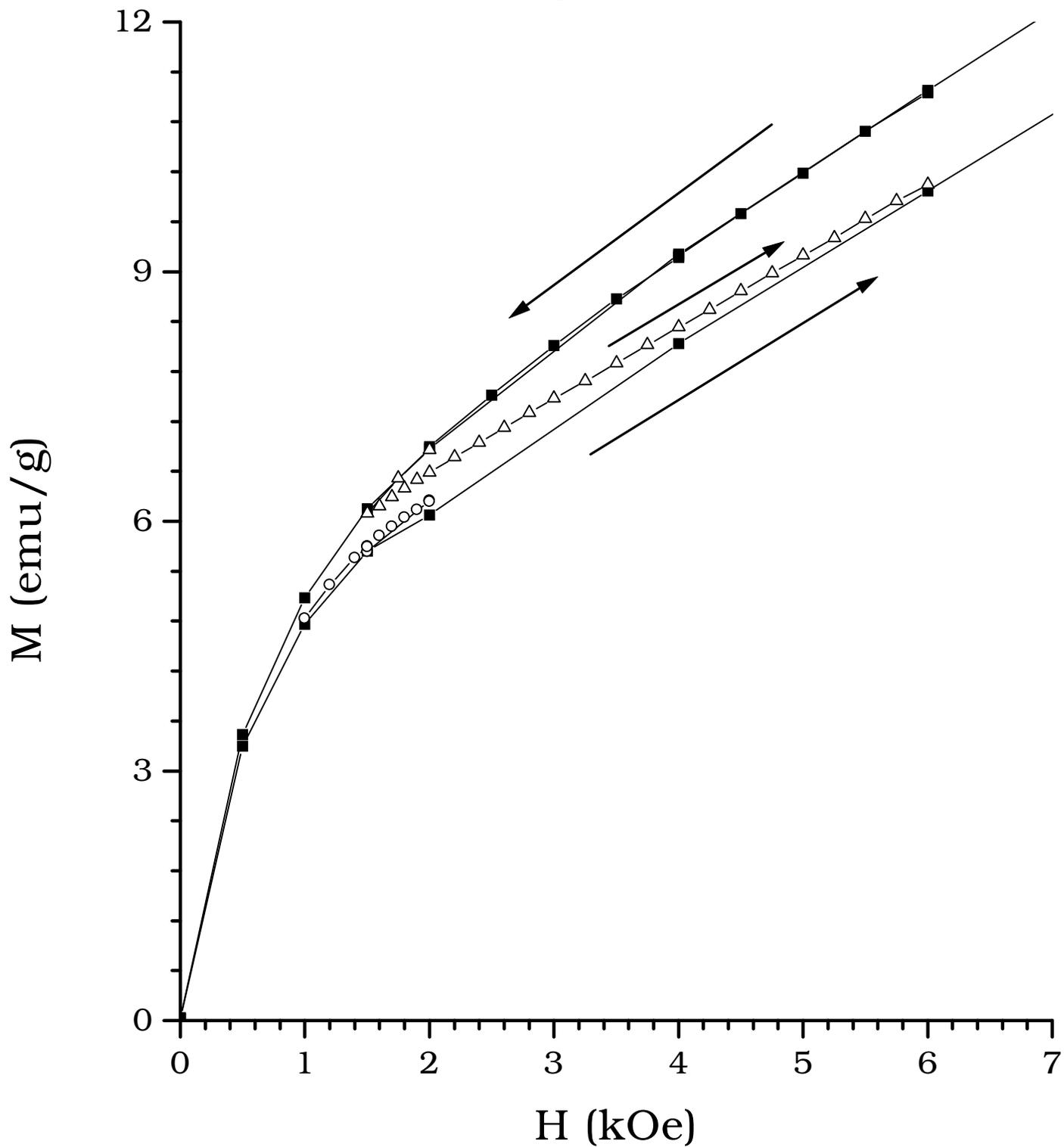
\end{document}